\documentclass[manuscript]{acmart}
\usepackage{todonotes}
\usepackage{amsmath}

\copyrightyear{2022}
\acmYear{2022}

\acmConference[FAccTRec '22]{Workshop on Responsible Recommendation}{September 18-23,
  2022}{Seattle, WA}
 \acmBooktitle{FAccTRec '22: Workshop on Responsible Recommendation, September 18-23,
  2022, Seattle, WA}
\setcopyright{none}

\begin{document}

\title{Who Pays? Personalization, Bossiness and the Cost of Fairness}

\author{Paresha Farastu}
\email{paresha.farastu@colorado.edu}
\affiliation{%
  \institution{University of Colorado, Boulder}
  \city{Boulder}
  \state{Colorado}
  \country{USA}
  \postcode{80027}
}

\orcid{}
\author{Nicholas Mattei}
\email{nmattei@tulane.edu}
\affiliation{%
  \institution{Tulane University}
  \city{New Orleans}
  \state{Louisiana}
  \country{USA}
  \postcode{}
}
\author{Robin Burke}
\email{robin.burke@colorado.edu}
\affiliation{%
  \institution{University of Colorado, Boulder}
  \city{Boulder}
  \state{Colorado}
  \country{USA}
  \postcode{80027}
}




\maketitle

\section{Introduction}
When designing a system to promote provider-side fairness \cite{burke_robin_multisided_2017}, implementers of recommender systems are primarily concerned with ensuring that protected group(s) of providers have a fair opportunity to promote their items or products through the recommendation opportunities (i.e., users) that the system provides. Typically, experiments have discovered that there is a ``cost of fairness'' that is borne by the consumer side of the interaction: users receive lower utility than they would in a system that is unconstrained by fairness objectives. This is to be expected: if the optimization of the system for user utility were perfect, then a more constrained optimization would not be able to achieve a better result.

The cost of fairness, borne by the consumer-side of the market, raises its own questions about fairness. In this paper, we look at the question of consumer-side fairness arising as a consequence of provider-side constraints. That is, \emph{are different users impacted differently by the imposition of the fairness constraint for providers?} For example, one might imagine that users who like mainstream, popular, items might get lower recall in their recommendations if the fairness constraint puts a broader range of content into everyone's lists. We refer to the loss of utility resulting from the imposition of a fairness constraint as the ``cost of fairness'' \cite{Goel:CostofFairness}.

One way that researchers have sought to lower the total cost of fairness is through a personalized approach \cite{liu2019personalized,mehrotra2018towards}. For example, the PFAR algorithm \cite{liu2019personalized} identifies users with diverse taste profiles as the ones more likely to be receptive to protected group items and weights the fairness objective more heavily when reranking for these users.
However, in adopting a personalized approach to the fairness objective, researchers may be opening their systems up to strategic behavior on the part of users. Users may deliberately manipulate their profiles to appear less flexible and, by doing so, force other users to bear the cost of fairness. This type of incentive has been studied in the computational social choice literature under the heading of ``bossiness.'' In this position paper, we explore an extension of this concept appropriate to the fairness-aware recommendation problem. 

\section{$\epsilon$-Bossiness}
``Nonbossiness, a criterion frequently used in the context of strategyproof allocation, ensures that individuals cannot be bossy, that is, change the assignment for others, by reporting different preferences, without changing their own \cite{papai2000strategyproof}.''  In other words, a desirable property of an allocation mechanism, i.e., one that allocates items to users, is that no user is able to strategically report (manipulate) their input to the system in a way that harm others without getting worse outcomes themselves. Many mechanisms are vulnerable to this type of strategic behavior. 

Within research on social choice including voting and allocation mechanisms, bossiness is what is called an axiom: either a mechanism (allocation rule, voting rule) is non-bossy, or it allows, for some set of inputs, some agent to be able to manipulate the system \cite{BCELP16a}. One of the most famous results related to bossiness is a characterization result: the only deterministic mechanism that is strategyproof, non-bossy, and neutral for the case of indivisible item allocation is random serial dictatorship, i.e., pick an agent at random, and let them pick their most desired item from a set \cite{svensson1999strategy}. While there are some studies of how often mechanisms that allow for bossiness admit such an action \cite{hosseini2018investigating}, we know of no work on relaxations of bossiness in settings that are more complex than one-to-one matchings.

In the fairness-aware recommendation context, an online, many-to-one matching setting, the user's input is their user profile, i.e., the set of ratings (or implicit actions) that they have taken that the system uses as evidence of interest or disinterest in particular items. Users may make strategic choices about their interactions with a platform in order to influence its profile of them \cite{ellison2020we}, for example, to avoid being targeted by related advertising. We assume for the following discussion that the recommender system learns, in some sense, a ``true'' representation of the user's preferences as given and uses this preference to provide recommendations that are in some sense optimal. 

Given this assumption on true preferences, the bossiness scenario from social choice does not arise because the user cannot alter their interactions away from their true preferences without the system's output also changing for the worse (assuming a singular optimal recommendation set). We therefore extend the concept of bossiness to cover a user's strategic manipulation of their input profile, even if it incurs some utility loss, in order to avoid a larger loss imposed by a constraint. Consistent with relaxations of other social choice concepts, we are terming this $\epsilon$-bossiness.

Formally, let $p_j$ be the profile associated with user $j$, and let $\ell_j$ be recommendation list or slate generated for user $j$ through some recommendation function $\mathcal{R}$, i.e., $ \mathcal{R}(p_j) \rightarrow \ell_j$. The list $\ell_j$ has some utility for user $j$, $ U(j, \ell_j) \rightarrow u_j $. If we replace $\mathcal{R}$ with a fairness-aware recommendation algorithm $\mathcal{F}$, there is a cost of fairness borne by each user ${\delta_{j}}^{f} = U(j, R(p_j)) - U(j, F(p_j)) \geq 0$, for all $j$. We will assume that the total amount of costs is fixed, determined by the system designers and some function of the fairness constraint on the provider side: $\Delta^f = \sum_j {\delta_{j}}^{f}$.\footnote{This might seem like a strong assumption; we consider alternate assumptions below.}

When $\mathcal{F}$ is personalized to each user, it will analyze each user's profile for indications about the degree of compatibility between the recommendation context and the user's interest and promote protected group items when they are most compatible. This kind of intervention has been found to enhance the fairness-accuracy tradeoff \cite{liu2019personalized,sonboli2020opportunistic} Inherently, this means that some users will bear less of the fairness cost. 

Suppose a bossy user $B$ presents to the system a profile that demonstrates inflexibility in order to take advantage of the personalized application of fairness. The system recognizes this as such and sticks closely to $B$'s preferences in generating recommendations, promoting protected group items in other users' lists, but not $B$'s. Hence, $B$ has (arguably) increased their utility (no ``cost of fairness'' for them) and decreased that of other users who now bear this cost, instead. 

This simple analysis leaves out however the fact that the user is building an artificially narrow profile on the system, which will may yield recommendations of lower utility. Let $p_a$ be the adversarial profile that the user adopts, such that the system imposes no fairness constraint on their results: $u_{a}^{f} = u_a$, meaning that $U(B, \mathcal{F}(p_a)) = U(B, \mathcal{R}(p_a))$. This does not mean that $u_{a}^{f} = u$, however, since we assume there is a cost of untruthfulness: $\delta_a = u_a - u$. But it is not necessarily the case that the cost of fairness is more negative. Let $\epsilon = \delta_a > \delta_f$. If $\epsilon > 0$ then $B$'s (bossy) strategy is effective and they have shifted some of the fairness cost to others.

One might argue that the assumption of fixed $\Delta^f$ is an overly strong one. There is no guarantee that the fairness cost not borne by one user is always taken up by another. However, if that is not the case and we assume that the fairness-aware algorithm is efficient, then less cost of fairness for the consumer side  means less fairness on the provider side. If our bossy user $B$ is not imposing an externality on fellow users, then they will be imposing it on the providers.

\section{What to do about it}

Strategies for avoiding giving an advantage to bossy users in this recommendation context come in several categories. We could drop the personalized approach to fairness. However, this may be hard in practice: attitudes w.r.t. protected group items may be part of the learned model. So, in some sense, fairness-aware recommendation will always be personalized, even if there is no explicit aspect of the algorithm that assesses user compatibility with a fairness concern. 

Another approach is to establish a constraint such that every user contributes a minimum amount towards the fairness objective. In practice, this looks something like the ``Rooney rule'' \cite{kleinberg2018selection}, specifying a minimum number of protected group items in every list. This strategy has nice theoretical properties, but could be alienating to users. If users note that their feeds always contain a less-relevant protected group item, they may learn to ignore them.

Another approach is to ignore the low end of the fairness cost distribution and concentrate instead on the high end. The approach used by \citet{patro2020fairrec} is to set a constraint on the maximum of fairness loss that can be incurred by an individual user. In this work, the bound is set through an envy-freeness (up to one item) $EF1$ constraint. A different mechanism is used by \citet{do2021two} where a Lorenz efficiency criterion is formulated as a max-min constraint in a quadratic programming solution for fair ranking. These approaches do not prevent an $\epsilon$-bossy user from lowering their cost of fairness, but it does mean that there is a limit to the effectiveness of the strategy. The non-bossy users cannot be forced to absorb too much of the cost from others. A similar approach is taken in other recent works, such as \cite{naghiaei2022cpfair,wu2021tfrom}.

An important point about the approaches in \cite{patro2020fairrec,do2021two,naghiaei2022cpfair,wu2021tfrom} is that these are batch optimization processes. The entire set of recommendations across all users is computed at once. This is the only way to achieve simultaneous balance across all the constraints. Recommendations in fielded sytems are of course delivered over time, online, and batch processing may not be an effective strategy for responding to real-world dynamics. An alternative to the batch optimization approach is to evaluate and respond to a system's fairness properties in real time as in \cite{sonboli2020and,burke2022multi}. We plan to study the vulnerability of such systems to bossy strategies in our future work.

\section{Related Work}
Within the game theory, social choice, and optimization literature there is a well established history of investigating the ``price of'' various metrics \cite{koutsoupias1999worst}. The price (cost) of fairness is the related concept which investigates the overall loss of system efficiency under a ``fairness'' constraint as measured against the unconstrained versions \cite{bertsimas2011price}. The price of fairness has been well studied in many social choice settings including voting \cite{goel2018relating,celis2018multiwinner} and allocation of scarce resources such as kidneys \cite{dickerson2014computational,dickerson2014price}. Within recommender systems research, there is related work on the cost of fairness in ride-hailing systems, for example, \cite{suhr2019two,cao2021optimization,shi2021learning} and others consider consumer fairness in terms of wait time. However, such systems do not consider personalized preferences on the consumer-side, so they are not susceptible to strategic manipulation. Finally, bossiness as we have described it bears some resemblance to \emph{shilling attacks} \cite{lam2004shilling} where an outside user injects users to change the recommendation output. However, note that this user is extrinsic to the system, and doesn't care about their recommendations, hence it is a fundamentally different problem.

\begin{acks}
Authors Farastu and Burke were supported by the National Science Foundation under grant IIS-2107577; author Mattei was supported by NSF Grant IIS-2107505.
\end{acks}

\bibliographystyle{ACM-Reference-Format}
\bibliography{scruff.bib}

\end{document}